# 7T MRI Synthesization from 3T Acquisitions

Qiming Cui[1,2][0009-0001-7993-3561], Duygu Tosun[1][0000-0001-8644-7724], Pratik Mukherjee[1][0000-0001-7473-7409], Reza Abbasi-Asl[1][0000-0001-7824-4628] *

[1] University of California, San Francisco, CA 94143, USA
[2] University of California, Berkeley, CA 94720, USA
*Reza.AbbasiAsl@ucsf.edu

**Abstract.** Supervised deep learning techniques can be used to generate synthetic 7T MRIs from 3T MRI inputs. This image enhancement process leverages the advantages of ultra-high-field MRI to improve the signal-to-noise and contrast-to-noise ratios of 3T acquisitions. In this paper, we introduce multiple novel 7T synthesization algorithms based on custom-designed variants of the V-Net convolutional neural network. We demonstrate that the V-Net based model has superior performance in enhancing both single-site and multi-site MRI datasets compared to the existing benchmark model. When trained on 3T-7T MRI pairs from 8 subjects with mild Traumatic Brain Injury (TBI), our model achieves state-of-the-art 7T synthesization performance. Compared to previous works, synthetic 7T images generated from our pipeline also display superior enhancement of pathological tissue. Additionally, we implement and test a data augmentation scheme for training models that are robust to variations in the input distribution. This allows synthetic 7T models to accommodate intra-scanner and inter-scanner variability in multisite datasets. On a harmonized dataset consisting of 18 3T-7T MRI pairs from two institutions, including both healthy subjects and those with mild TBI, our model maintains its performance and can generalize to 3T MRI inputs with lower resolution. Our findings demonstrate the promise of V-Net based models for MRI enhancement and offer a preliminary probe into improving the generalizability of synthetic 7T models with data augmentation.

## 1 Introduction

There has been growing interest in testing the clinical advantages of Ultra-high-field Magnetic Resonance Imaging (MRI) at 7 Tesla (T) over high-field MRI at 3T. Many studies have shown that 7T MRIs have advantages in delineating anatomical structures that are relevant for identifying and monitoring pathological tissue [14]. For example, compared to 3T MRIs, 7T acquisitions can lead to better parcellations of subcortical structure and increase lesion conspicuity [20]. Recent studies have demonstrated the promise of 7T MRIs in assessing neurological disorders such as Epilepsy [26], Multiple Sclerosis [22], Parkinson's disease [27], Alzheimer's disease [23], and Traumatic brain injury [11]. For context, a visual comparison of lesion tissues in a mild TBI patient imaged across different field strengths is given in Fig. 1, which illustrates how 7T acquisition can be superior for delineating pathological features. While there is clinical value behind 7T neuroimaging, the availability of 7T MRI scanners is limited. Currently there are 109 reported 7T MRI scanners in the world, and they are concentrated in developed nations [10].

To make the clinical advantage of 7T neuroimaging more accessible, there have been attempts to construct algorithms that can enhance 3T MRIs to a 7T-like state by leveraging paired 3T and 7T MRIs [2, 3, 7, 13, 24, 29]. These studies have primarily focused on T1-weighted (T1w) MRIs [2, 3, 24] and Diffusion/Diffusion-weighted MRIs [5, 7, 13]. Other studies have attempted similar transformation tasks for MRIs acquired at lower field strengths, such as enhancing 0.36T MRIs to 1.5T/3T MRIs [17]. Specifically, the WATNet [24] model currently achieves state-of-the-art performance in synthesizing T1w 7T MRI. This model is based on an encoder-decoder 2D convolutional neural network fused with wavelet-based feature injection. We have selected WATNet as the benchmark for making performance comparisons in this study.

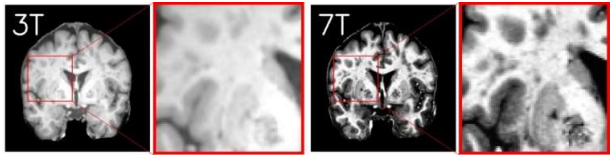

**Fig. 1. Comparison of T1w MRIs from a TBI patient imaged at 3T (left) and 7T(right).** White matter lesions and subcortical microbleeds are more visible in the 7T scan.

Current studies on synthetic 7T generation do not present evaluations based on pathological tissue features. This is limiting for assessing the clinical value of synthetic 7T because the advantage of 7T imaging lies in its depiction of pathological tissue. As such, we believe the construction and assessment of a synthetic-7T model is best conducted in a setting where a discernible amount of pathological tissue is present. To address this limitation, we perform evaluations on a 3T-7T dataset consisting of patients diagnosed with mild traumatic brain injury (TBI) in conjunction with a public dataset consisting of healthy subjects. Since TBI leads to visible pathological features such as white matter lesions and traumatic microbleeds [18], studying synthetic 7T generation with a TBI dataset offers a novel look on the clinical value of synthetic 7T images.

Input heterogeneity has also not been sufficiently explored in the context of synthetic 7T generation. Due to differences in acquisition protocols and unwanted inter-scanner variabilities across data collection sites [28], deep learning models intended for MRI applications should have a degree of generalizability to input variation. To address this, we implemented a custom data augmentation scheme to construct a synthetic-7T model that, based on our assessment, is robust to degradation of the input MRI. This data augmentation can enable future studies on larger multi-site datasets.

Our main contributions in this study are three folds: (1) Multiple novel variants of V-Net model for 3T-to-7T enhancement of T1w MRIs, leading to state-of-the-art performance. (2) The first qualitative evaluation of pathological tissue enhancement via synthetic 7T generation. (3) Demonstrating model generalizability to low-resolution MRIs by utilizing a multisite dataset with a custom data augmentation scheme.

Our code for this study is freely available at:
**https://github.com/abbasilab/Synthetic_7T_MRI**

## 2  Method

We designed and trained three V-Net based models and one WATNet model for synthetic-7T generation. All models were implemented in PyTorch and trained with 2 Nvidia 4090 GPUs. Both the WATNet model and the V-Net model were trained with mean absolute error (MAE) as the objective function. MAE loss is preferable to mean squared error (MSE) for 3T-to-7T image enhancement, since 7T images in our dataset contain stronger vascular pulsation artifacts. In addition, there has been evidence that 7T MRIs are more prone to other forms of artifacts [15]. These artifacts manifest as signal hyperintensities that will heavily influence the loss if MSE is used as the objective function instead of MAE. For a given 3T image, $I_{3T}$, a ground truth 7T image, $I_{7T}$, and a synthetic-7T model, G, the MAE loss is defined as:

$$L_{MAE} = \frac{1}{n}\sum_{i=1}^{n}|I_{7T} - G(I_{3T})| \quad (1)$$

**V-Net.** We implemented a 5-layer V-Net Model [19] with a modified decoder branch where we changed the transposed convolutions to nearest neighbor upsampling. The V-Net model utilizes 3D convolutions, with skip connections across layers in the encoder and decoder branches. The V-Net model is trained with the MAE loss. We utilize nearest neighbor upsampling instead of transposed convolutions for deconvolution since using the latter often results in "checkerboard" artifacts in the predicted images [21].

**Perceptual V-Net.** We implemented a perceptual V-Net by extracting perceptual loss from the encoding branch of the SynthSeg model. The perceptual V-Net is similar to the SRResNet in the SRGAN paper [16], where a baseline deep learning model is augmented with a perceptual loss for enhancing the perceptual quality of images. The original SRGAN design used the VGG network for extracting perceptual loss. However, the VGG model was trained for natural image classification rather than a task adjacent to medical imaging. Thus, we hypothesize that the hidden state representation of SynthSeg is a more robust feature extractor for brain MRIs, since SynthSeg is a deep learning model trained to generate brain segmentation maps in a contrast-agnostic and resolution-agnostic fashion [4]. We define SynthSeg loss as the output of the 3$^{rd}$ layer in the encoding branch of the SynthSeg Model. Given a synthetic-7T generation model, G, an input 3T image, $I_{3T}$, and a 7T image, $I_{7T}$, SynthSeg loss is defined as:

$$L_{SynthSeg} = \frac{1}{n}\sum_{i=1}^{n}|SynthSeg_{layer3}(I_{7T}) - SynthSeg_{layer3}(G(I_{3T}))| \quad (2)$$

The sum of $L_{MAE}$ and $L_{SynthSeg}$ is used for training the perceptual V-Net.

**V-Net-GAN.** We also constructed a generative adversarial V-Net model for synthetic 7T generation. The principle of V-Net-GAN's design is similar to SRGAN in the use of perceptual loss functions, but the two models have different architectural choices. The generator architecture of V-Net GAN is a V-Net, and the discriminator is a "half" V-Net truncated at the bottleneck. The output of the truncated V-Net is passed through a global sum pooling layer, a ReLu layer, and a linear layer for the discriminator's decision output. We also use the Wasserstein GAN loss for more stable model convergence [1]. Given a discriminator model, D, and a generator model, G, the adversarial loss is defined as:

$$L_{adv} = -\left(\frac{1}{n}\sum_{i=1}^{n} D(I_{7T}) - \frac{1}{n}\sum_{i=1}^{n} D\bigl(G(I_{3T})\bigr)\right) \quad (3)$$

The sum of $L_{MAE}$, $L_{\text{SynthSeg}}$, $L_{adv}$ is used for training the V-Net-GAN.

**WATNet.** We used the WATNet architecture [24] as the current benchmark for 7T synthesization. The WATNet model is a 2D convolutional model that utilizes wavelet-based feature extraction. The wavelet features are injected into the encoder branch of the model to allow for fusion of information across the spatial and wavelet domains. We re-implemented the WATNet model in PyTorch to enable training with more recent GPUs. It is worth noting that in the original implementation of WATNet in Caffe, the convolution operation is designed in 2D only. To incorporate the 3D spatial information, the original WATNet implementation used 3 adjacent MRI slices as 3 separate input channels. We replicated this design choice for an accurate re-implementation of the original model. WATNet is trained with the MAE loss.

## 3     Experimental Design

**Dataset.** Our primary dataset contains imaging data from subjects (n=8) diagnosed with mild TBI. The dataset was collected at the University of California, San Francisco. This dataset consists of T1w MPRAGE images with an isotropic resolution of 0.8mm collected at a 3T Siemens Skyra scanner and T1w MP2RAGE images with an isotropic resolution of 0.7mm collected at a 7T Siemens Magnetom scanner. In addition to this dataset, we also assembled a multisite dataset (n=18) by harmonizing the primary TBI dataset with a publicly available dataset of paired 3T-7T MRIs collected from healthy subjects (n=10) [6]. This second dataset contains T1w MPRAGE images with an isotropic resolution of 0.8mm collected at a 3T Siemens Magnetom Prisma scanner, and T1w MP2RAGE images with an isotropic resolution of 0.65mm collected at a 7T Siemens Magnetom scanner. To account for inter-scanner variability we used the RAVEL [8] harmonization algorithm for both intensity normalization and removal of undesirable inter-subject technical variabilities across the two datasets. During preprocessing, all images were aligned to a 0.7mm MNI-152 template, and each 3T image was aligned to their 7T counterpart with affine transformation using the FSL FLIRT toolbox [12].

**Data augmentation.** To construct a model with greater generalizability, we implemented a data augmentation scheme aimed at simulating a wider range of input distributions. We applied data augmentation when training with the harmonized dataset. Prior to training, every image pair in the training set was augmented with two transformed 3T-7T pairs based on the following transformations: (1) random flip along the coronal or sagittal plane. (2) -20 to 20 degree rotation in any direction (3) random elastic deformation. Additionally, we applied the following transformation to 3T images only: (1) Gamma correction with a random γ value ranging from $e^{-0.3}$ to $e^{0.3}$. (2) Resolution downsampling with a scale factor ranging from 1 to 5, followed by resampling to the original resolution. We appended the augmented image pairs to the original training set, resulting in 45 training image pairs for each fold in the 6-fold cross validation analysis.

**Training details.** Images were broken down into patches during training. For models based on the V-Net architecture, the patch size was 64x64x64, with a batch size of 40. For the WATNet model, the patch size was 64x64x3, with a batch size of 128. Models were trained using the ADAM optimizer. When training without data augmentation, all models were trained for 300 epochs. When training with data augmentation, all models were trained for 500 epochs. We used a learning rate of 1e-3 for the V-Net models and a learning rate of 1e-4 for the WATNet model.

**Evaluation.** We performed quantitative evaluation of the model with peak signal-to-noise ratio (PSNR) and structural similarity index measure (SSIM). For the models trained with data augmentation, we also generated synthetic-7T images from downsampled inputs. This is aimed at simulating 3T acquisitions with lower quality. Additionally, we use the SynthSeg model to generate multi-class brain segmentations for synthetic and ground truth 7T images. We calculate the multiclass dice coefficient as an alternative metric to gauge the quality of synthetic 7T images.

For comparing performance across all four model architectures, we train each model on the TBI dataset and perform leave-one-out cross validation. Additionally, we assess the WATNet and V-Net model's ability for data generalization by training them on the harmonized dataset with our data augmentation scheme. Due to limited compute capacity, we performed 6-fold cross validation for the second evaluation. All quantitative performance metrics in this study are reported based on the cross-validation analyses. For qualitative assessment, we used final models trained on all available data for generating image predictions.

## 4  Results

### 4.1  Our V-Net based model for 7T MRI synthesization archives state-of-the-art performance

The V-Net based model (SSIM: 0.914±0.016, PSNR: 25.60±0.77) outperforms the benchmark WATNet model (SSIM: 0.881±0.008, PSNR: 23.60±0.37) in all three performance metrics (Fig. 2). The WATNet prediction has an SSIM value that closely approximates the WATNet performance reported in [24] (SSIM of 0.878).

The performance metric of the perceptual V-Net model with SynthSeg loss (V-Net-SSeg) (SSIM: 0.908±0.011, PSNR: 25.51±0.74) is lower than V-Net, and the performance metric of the V-Net-GAN (SSIM: 0.899±0.012, PSNR: 24.33±0.93) is lower than both the V-Net and V-Net-SSeg, but both models outperform the WATNet model. The performance drop of V-Net-SSeg and V-Net-GAN compared to the V-Net model is expected considering similar performance drops seen in natural image super-resolution with SRGAN, where improvements in perceptual quality can lead to hallucinations that decrease the quantitative measures of similarity to the ground truth [16]. Multiclass dice coefficients calculated based on SynthSeg brain segmentation indicate that V-Net and V-Net-SSeg have comparable segmentation performance (**Fig. 2C**). This suggests that the enhancement of tissue boundaries may be comparable across the two model variants.

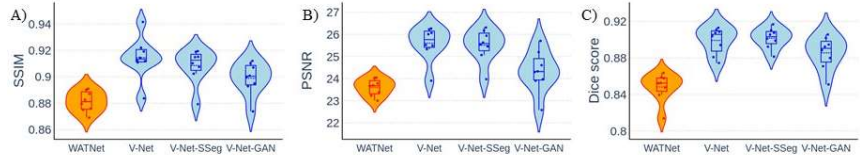

**Fig. 2. Violin plots of the quantitative performance metrics across models.** Individual datapoints represent model performance from one cross-validation fold. A) SSIM metrics. B) PSNR metrics. C) Multi-class Dice scores between segmentations generated from synthetic 7Ts and natural 7Ts.

### 4.2    Qualitative evaluation of pathological tissue enhancement via synthetic 7T generation

We provide a qualitative comparison of model outputs for patients with mild TBI to investigate how well synthetic-7T models enhance pathological tissue (**Fig. 3,** additional examples in **Supplemental Materials**). In **Fig. 3**, we selected an example region with white matter lesions and microbleeds in subcortical areas. In this case, there are two qualitative features of interest. First, the conspicuity of pathological tissue is higher in 7T images. This is evident in the separation of adjacent lesions and the sharper contour of subcortical microbleeds. Additionally, the 7T image better captures the heterogeneity within white matter lesions. This heterogeneity can be a useful clinical signal in neurodegenerative disorders such as multiple sclerosis. For example, previous studies have found that the heterogeneity within lesions can be used for elucidating lesion structure and improving diagnostic accuracy [25]. In the example presented in **Fig. 3**, it is evident that all V-Net based architectures outperform the WATNet for capturing both tissue conspicuity and lesion heterogeneity. The differences across the three V-Net-based models are more nuanced, but in this case, the V-Net-GAN model performs better for capturing smaller structural details such as the lesion feature highlighted in blue and the shape of microbleeds highlighted in cyan. Additional qualitative comparisons of model predictions are presented in the **Supplemental Materials**.

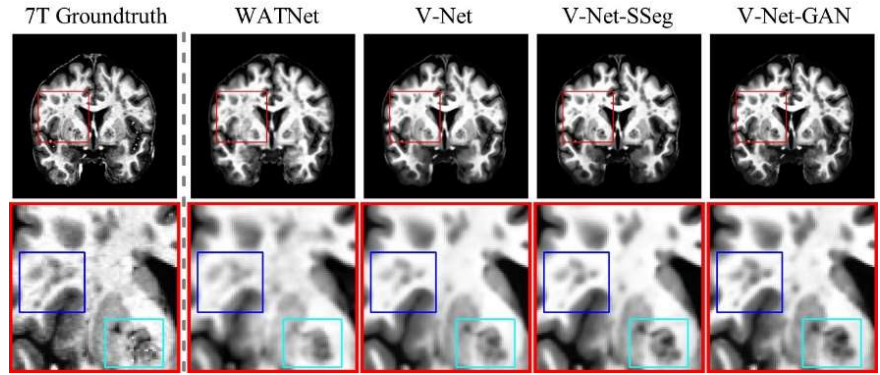

**Fig. 3. Qualitative comparison of synthetic 7T images in a subject with mild TBI.** The bounding box highlights white matter lesions (in blue) and subcortical microbleeds (in cyan).

### 4.3 Model generalizability to low-resolution MRI input

To test the ability of models to generalize to different input distributions, we also trained the WATNet and V-Net models on a harmonized dataset (n=18, both TBI and healthy subjects) in combination with data augmentation. In this training scenario, both architectures are capable of generalizing to downsampled inputs without significant performance drops (**Fig. 4A & 4B**, detailed quantitative values are presented in **Supplemental Materials**). However, V-Net retains a qualitative advantage over WATNet and enhances pathology with more fidelity when generalizing to lower input resolutions. This is evident in the qualitative comparison based on a TBI subject (**Fig. 5**, additional examples in **Supplemental Materials**). In this example, the WATNet prediction blurs the boundary of the lesion tissue even when the input is at its original resolution. This effect is exacerbated when the input is downsampled by a factor of two, and the synthetic 7T image displays a distortion of the lesion tissue highlighted in blue. In contrast, the V-Net model generates synthetic 7T images with high lesion conspicuity both at the original input resolution and at a downsampled resolution.

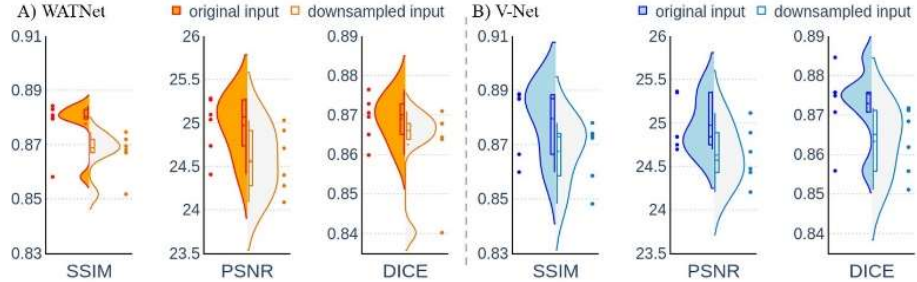

**Fig. 4. Violin plots comparing synthetic 7T performance between original and downsampled inputs.** Individual datapoints represent model performance from one cross-validation fold. Both models are trained on the harmonized dataset with data augmentation. A) WATNet metrics. B) V-Net metrics.

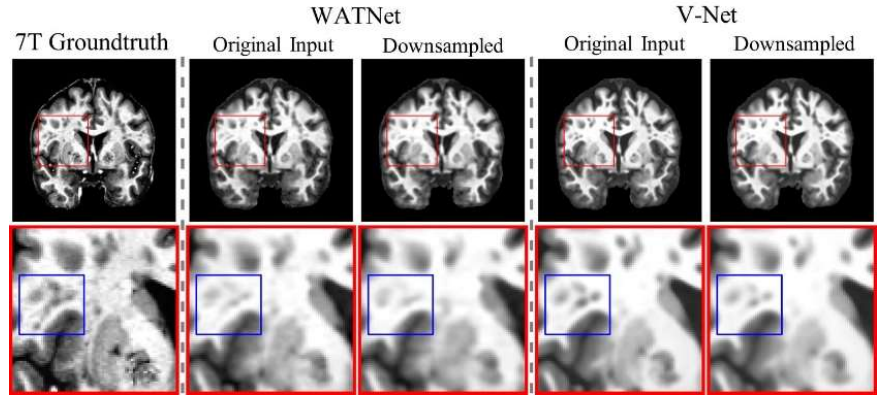

**Fig. 5. Comparison of model predictions from a TBI patient.** Input image resolution is downsampled by a factor of 2 in the downsampled scenario.

## 5    Conclusion and Future work

In this work, we presented a V-Net based algorithm for enhancing 3T MRIs by synthesizing 7T acquisitions that approximate natural 7T MRIs. On a dataset of paired 3T-7T MRIs from patients with mild TBI, our method achieves a superior performance compared to WATNet, the current state-of-the-art architecture for synthetic 7T generation. Additionally, our model enhances pathological tissue with more fidelity for clinically relevant insights. Our qualitative assessment on pathological tissue enhancement is a new step towards evaluating clinical applications of synthetic 7T models.

While synthesization techniques based on machine learning frameworks demonstrate remarkable performance, their application in clinical settings require extensive validation. Our qualitative assessment of the generated images is the first step towards this goal, but future work should include formal quantification of model hallucination and uncertainty. This is possible through extensive clinical assessment of model findings and clinical rating of model-generated images.

Our models in this study were trained on T1w images for synthetic 7T generation, however, clinical use cases often involve viewing multiple MR contrasts in conjunction. Overcoming this limitation requires collecting 3T-7T data pairs at multiple contrasts which should be the focus of future data collection efforts. Another potential challenge in 7T MRI synthesization from 3T acquisitions is the accurate alignment of 3T images to corresponding 7T ones. This is primarily due to the structural differences and varying degrees of artifacts between 3T and 7T MRIs. Future studies on deep-learning-based registration methods (e.g., [9]) that are more contrast agnostic and robust to structural variations could lead to more robust synthetic 7T generation pipelines.

**Acknowledgements.** The authors would like to thank the UCSF DeepNeuro team for their comments. RA would like to acknowledge funding from the Weill Neurohub.

**Disclosure of Interests.** The authors have no competing interests to declare that are relevant to this article.

# Supplemental Materials

**Table S1.** Leave-One-Out cross validation performance metrics across synthetic 7T models.

| Task | WATNet | V-Net | V-Net-SSeg | V-Net-GAN |
|---|---|---|---|---|
| SSIM | *0.881±0.008* | *0.914±0.016* | *0.908±0.013* | *0.899±0.012* |
| PSNR | *23.6±0.37* | *25.60±0.77* | *25.51±0.74* | *24.33±0.93* |
| DICE | *0.848±0.016* | *0.899±0.015* | *0.902±0.011* | *0.886±0.018* |

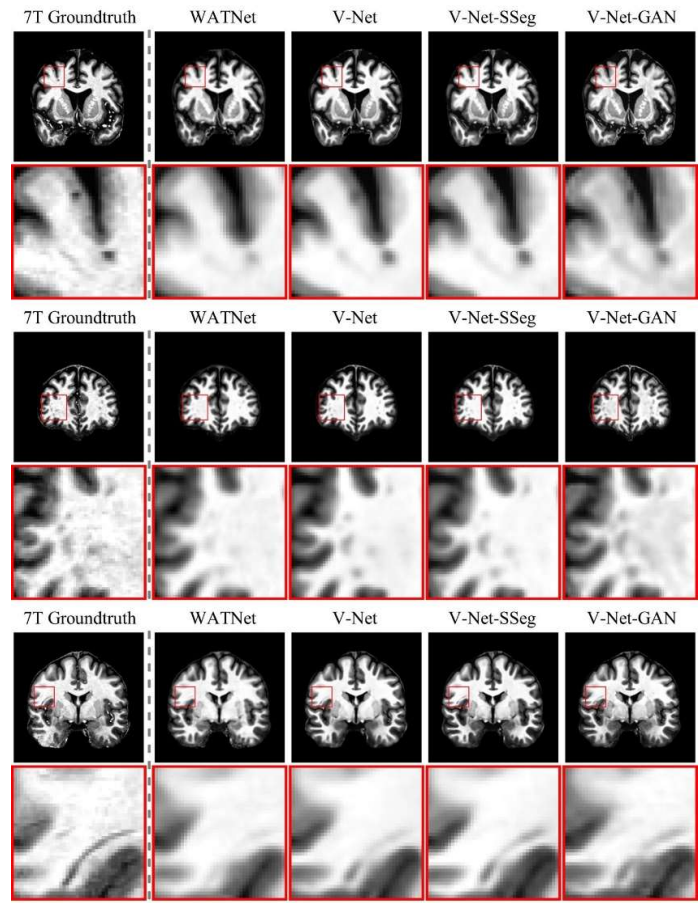

**Fig. S1. Qualitative comparison of synthetic 7T images across three subjects with mild TBI.** Every row corresponds to a unique subject.

**Table S2.** 6-fold cross validation results for WATNet trained with data augmentation.

| Task | WATNet SSIM | WATNet PSNR | WATNet DICE |
|---|---|---|---|
| Original resolution | 0.877±0.010 | 24.97±0.34 | 0.869±0.006 |
| Downsampled (s=2) | 0.867±0.008 | 24.57±0.37 | 0.862±0.011 |

**Table S3.** 6-fold cross validation results for V-Net trained with data augmentation.

| Task | V-Net SSIM | V-Net PSNR | V-Net DICE |
|---|---|---|---|
| Original resolution | 0.880±0.013 | 24.97±0.30 | 0.873±0.010 |
| Downsampled (s=2) | 0.867±0.012 | 24.63±0.33 | 0.863±0.009 |

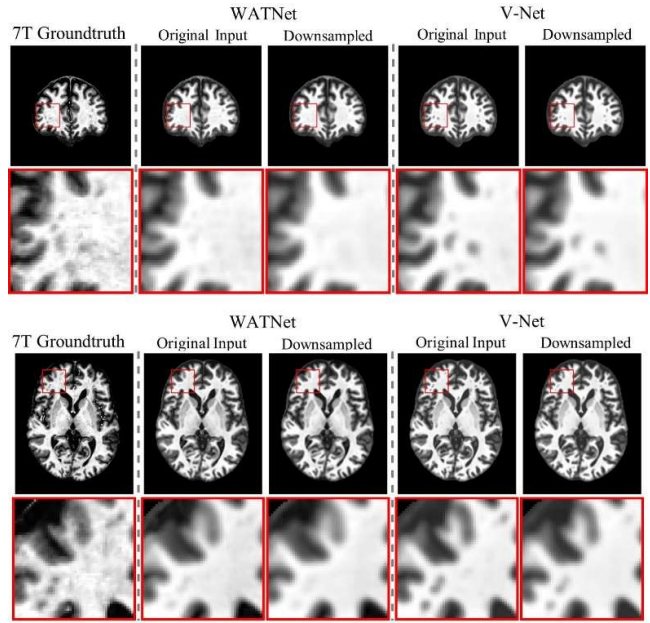

**Fig. S2. Comparison of synthetic 7T images generated from original and downsampled input MRIs.** Every row corresponds to a unique subject. Input image resolution is downsampled by a factor of 2 in the downsampled scenario.